\documentclass{article}
\usepackage{spconf,amsmath,graphicx}
\usepackage{amsmath,amssymb,amsfonts,mathrsfs}
\usepackage{algorithm,algorithmic}
\usepackage{booktabs}
\usepackage{multirow}
\usepackage{color}
\usepackage{caption} 
\usepackage{etoolbox,siunitx}
\usepackage{kantlipsum}
\usepackage{url}
\usepackage{fancyhdr}

\usepackage{setspace}

\renewenvironment{thebibliography}[1]{
  \begin{oldthebibliography}{#1}
}
{
  \end{oldthebibliography}
}
\usepackage[
backend=biber,
style=ieee,
citestyle=numeric-comp,
maxbibnames=3,
maxcitenames=3,
doi=false,isbn=false,url=false,eprint=false
]{biblatex}
\defbibheading{bibliography}[\refname]{}

\newcommand{\modify}[1]{{\color{black}#1}}

\title{End-to-End Integration of Speech Recognition, Dereverberation, Beamforming, and Self-Supervised Learning Representation}
\name{Yoshiki Masuyama$^{1}$, Xuankai Chang$^{2}$, Samuele Cornell$^{3}$, Shinji Watanabe$^{2}$, Nobutaka Ono$^{1}$}
\address{
$^1$Tokyo Metropolitan University, Japan \,\,
$^2$Carnegie Mellon University, USA \\
$^3$Università Politecnica delle Marche, Italy
}
\begin{document}
\ninept
\maketitle
\begin{abstract}
Self-supervised learning representation (SSLR) has demonstrated its significant effectiveness in automatic speech recognition (ASR), mainly with clean speech.
Recent work pointed out the strength of integrating SSLR with single-channel speech enhancement for ASR in noisy environments.
This paper further advances this integration by dealing with multi-channel input. We propose a novel end-to-end architecture by integrating dereverberation, beamforming, SSLR, and ASR within a single neural network.
Our system achieves the best performance reported in the literature on the CHiME-4 6-channel track with a word error rate (WER) of $1.77\%$.
While the WavLM-based strong SSLR demonstrates promising results by itself, the end-to-end integration with the weighted power minimization distortionless response beamformer, which simultaneously performs dereverberation and denoising, improves WER significantly.
Its effectiveness is also validated on the REVERB dataset.
\end{abstract}
%

\section{Introduction}
\label{sec:intro}

The progress of deep learning has significantly improved the performance of automatic speech recognition (ASR)~\cite{Hinton2012,Chiu2018}.
Recently, the end-to-end (E2E) framework has achieved promising results and has become popular owing to its simplicity.
In the literature, various sequence-to-sequence modeling techniques have been developed, such as connectionist temporal classification
(CTC)~\cite{Graves2006,Miao2015}, attention-based encoder--decoder~\cite{Chorowski2015,Chan2016}, and the recurrent neural network transducer~\cite{Graves2012}.
Neural network (NN) architectures have also been investigated, including Transformer~\cite{Karita2019} and Conformer~\cite{Gulati2020}.
Although these developments have improved the ASR performance under clean conditions, ASR in noisy reverberant environments is still a challenging problem.
This paper addresses this problem by combining E2E ASR with self-supervised learning representation (SSLR) and multi-channel speech enhancement (SE).

SSL aims to obtain a good data representation by solving a pretext task~\cite{Mohamed2022}.
As the pretext task is defined without manual labels, SSL can leverage a large amount of unlabeled data.
The generalization capability of the representation has been confirmed in various speech processing tasks~\cite{Yang2021,Tsai2022}.
Recently, the hidden unit bidirectional encoder representations from Transformers (HuBERT) demonstrated great potential in ASR~\cite{Hsu2021}.
HuBERT trains NN to predict the $k$-means cluster of the speech feature on the masked region.
Most SSL methods use only clean speech, and thus their strength is limited under noisy reverberant conditions~\cite{Chang2021}.
Recently, SSL using both clean and noisy speech has been investigated to obtain robust representations~\cite{Chen2021,Wang2022,Wang2022a,Huang2022}.
A variant of HuBERT, called WavLM~\cite{Chen2021}, achieved state-of-the-art performance in SUPERB benchmark tasks~\cite{Yang2021} by predicting the cluster of clean speech from noisy and overlapped speech.

\begin{figure}[t]
\centering
\includegraphics[width=0.96\columnwidth]{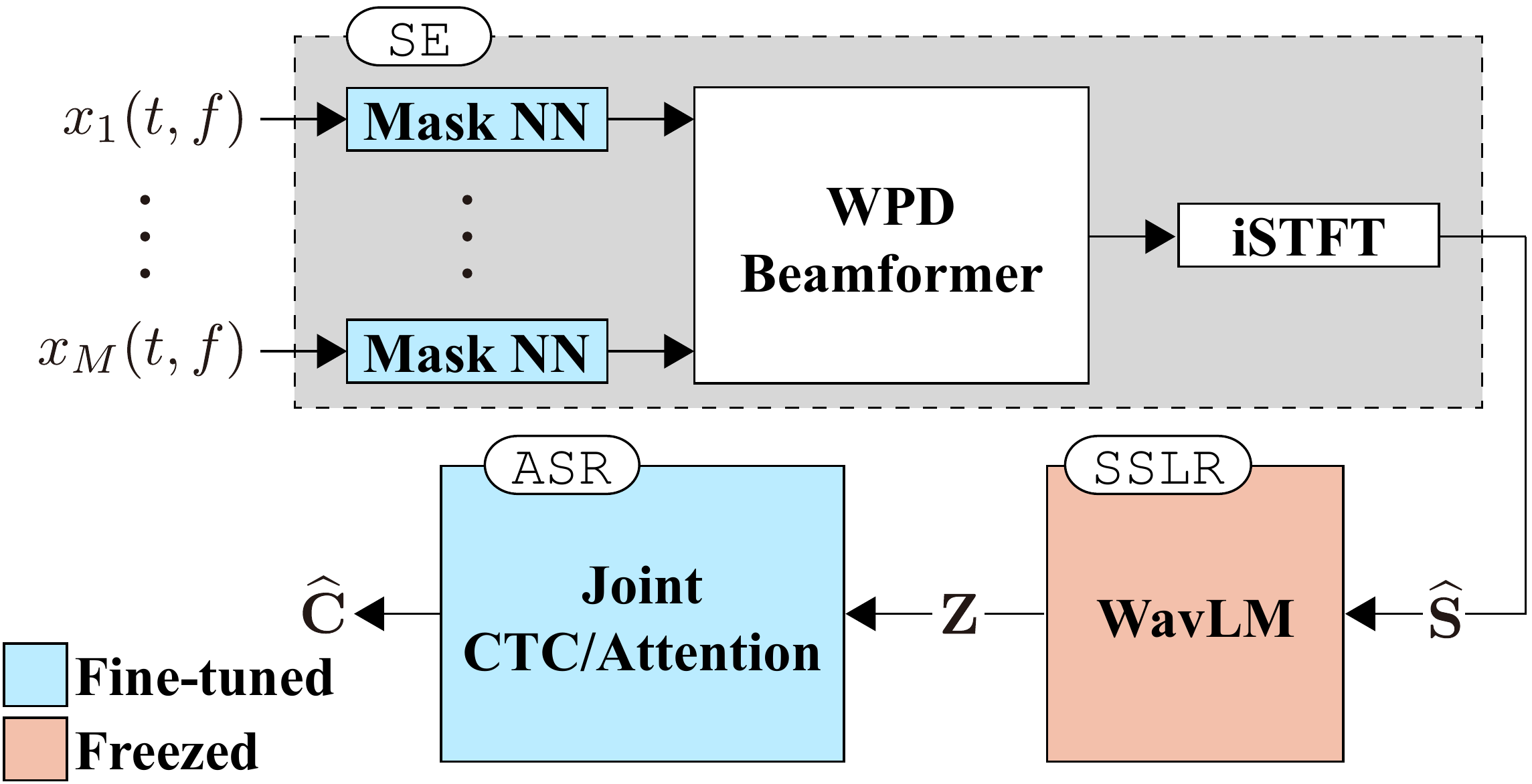}
\caption{Block diagram of MultiIRIS.
Colored blocks have learnable parameters.
In the joint training, we fine-tune the mask NN and the \modify{joint} CTC/attention-based encoder--decoder while WavLM parameters are not updated.}
\label{fig:overview}
\vspace{-0.3cm}
\end{figure}

In this paper, we propose to integrate multi-channel SE, SSLR, and E2E ASR into a single NN that is jointly optimized based on an ASR criterion.
As illustrated in Fig.~\ref{fig:overview}, we combine the weighted power minimization distortionless response (WPD) beamformer, which performs dereverberation and denoising simultaneously, and WavLM with E2E ASR.
While the WPD beamformer can leverage the spatial diversity of a given multi-channel noisy signal, WavLM can exploit a large amount of unlabeled single-channel data.
We expect them to complement each other.
In a relevant work~\cite{Chang2022}, a similar system for the single-channel case was named the integrated speech recognition with enhanced speech input for self-supervised learning representation (IRIS).
Building upon this work, we develop novel multi-channel end-to-end integration and name our system multi-channel IRIS (MultiIRIS).
Through experiments, we confirmed the effectiveness of combining multi-channel SE and robust SSLR.
MultiIRIS achieved state-of-the-art performances on the CHiME-4~\cite{Vincent2017} and REVERB~\cite{Kinoshita2016} datasets.

\vspace{-0.2cm}
\section{Related Work}

\subsection{Multi-channel SE for ASR}
\label{sec:mcse_for_asr}

Multi-channel SE, including dereverberation and beamforming, has been successfully used for robust ASR.
For dereverberation, the weighted prediction error (WPE)~\cite{Nakatani2010} contributed to obtaining the best results in several challenges~\cite{Kinoshita2016,Barker2018}.
When multiple microphones are available, beamforming can efficiently suppress unwanted interferers and reduce reverberation.
In the literature, the minimum variance distortionless response (MVDR) and minimum power distortionless response (MPDR) beamformers have been widely used~\cite{Souden2010}.
Recently, their unified version, called the WPD beamformer~\cite{Nakatani2019}, has been proposed for simultaneous dereverberation and denoising, and it outperformed the cascaded system of WPE and the MPDR beamformer.
One drawback of the original WPD beamformer is that it requires an iterative optimization procedure to obtain the filter.
To address this problem, a recent work proposed to use NN and estimate this filter in a closed form~\cite{Zhang2020a}.
This enables us to optimize NN based on the output of the WPD beamformer without iterations \cite{Zhang2020a,Ni2021}.

\vspace{-4pt}
\subsection{Joint Training of SE and ASR}

Joint training of SE and ASR models has gained increasing attention~\cite{Heymann2017,Ochiai2017}.
When both models are differentiable, the SE model can be optimized through backpropagation with an ASR criterion together with the ASR model.
This joint training has various advantages.
First, the SE model is optimized in terms of the final criterion instead of an enhancement criterion that is not guaranteed to be correlated with ASR performance.
Second, this training does not require pairs of noisy and clean signals that are difficult to obtain without a careful recording setup or simulation.
As a result, we can leverage real recordings with transcripts to adapt an SE model pre-trained on simulated data to real-world data.
Owing to these advantages, joint training has been investigated not only for denoising but also for dereverberation~\cite{Subramanian2019} and speaker separation~\cite{Chang2019}.
While SSLR was not used in these studies, IRIS~\cite{Chang2022} achieved state-of-the-art performance on the CHiME4 single-channel track by integrating an SSLR model between single-channel SE and E2E ASR models.
While IRIS is based on a fully-neural SE method, MultiIRIS leverages well-studied beamforming techniques supported by NN.
We expect that this will allow to exploit the spatial diversity of a multi-channel signal and make our system more robust.

Robust ASR without explicit SE models has also been investigated~\cite{Sainath2015,Shao2022}.
Although these methods also achieved promising results, they are often not easy to interpret and require much more training data.
Our end-to-end integration maintains the modularity of multi-channel SE and ASR, which makes the system more interpretable.
In addition, the modularity enables us to pre-train the ASR model on massive amounts of single-channel data.

\section{Proposed Multi-channel IRIS}
\label{sec:multiiris_description}

MultiIRIS consists of three models: multi-channel SE (\texttt{SE}), single-channel SSLR (\texttt{SSLR}), and E2E ASR (\texttt{ASR}), as illustrated in Fig.~\ref{fig:overview}:
\begin{align}
    \widehat{\mathbf{S}} &= \texttt{SE} (\mathbf{X}), \label{eq:speech-enhancement} \\
    \mathbf{Z} &= \texttt{SSLR} (\widehat{\mathbf{S}}), \label{eq:feature-extraction} \\
    \widehat{\mathbf{C}} &= \texttt{ASR} (\mathbf{Z}),
\end{align}
where $\mathbf{X}$ is the observed multi-channel noisy signal, and $\mathbf{S}$ and $\mathbf{C}$ are the target source image and its transcription, respectively.
Here, $\widehat{(\cdot)}$ denotes the estimate of its input.
In \eqref{eq:feature-extraction}, the \texttt{SSLR} model transforms the enhanced signal to features $\mathbf{Z}$.

\vspace{-4pt}
\subsection{Joint Dereverberation and Denoising by WPD Beamformer}

Let us denote a noisy signal observed by $M$ microphones
as $\mathbf{x}(t,f) \modify{\triangleq} [x_1(t,f), \ldots, x_M(t,f)]^\mathsf{T} \in \mathbb{C}^M$, where $t$ and $f$ are respectively the time and frequency indices, and $(\cdot)^\mathsf{T}$ denotes the transpose.
In the short-time Fourier transform (STFT) domain, we model the multi-channel noisy signal as:
\begin{equation}
    \mathbf{x}(t,f) = \mathbf{s}(t,f) + \mathbf{r}(t,f) + \mathbf{n}(t,f),
\end{equation}
where $\mathbf{s}(t,f) \in \mathbb{C}^M$ is the target source image, $\mathbf{r}(t,f) \in \mathbb{C}^M$ is the late reverberation, and $\mathbf{n}(t,f) \in \mathbb{C}^M$ is the noise.
To estimate $\mathbf{s}(t,f)$ from $\mathbf{x}(t,f)$, we further assume that the source image and the late reverberation can be modeled as 
\begin{align}
    \mathbf{s}(t,f) &= \mathbf{a}(0, f) s_1(t,f), \label{eq:instantaneous-model} \\
    \mathbf{r}(t,f) &= \sum_{\tau=\delta}^{\delta+\Delta\modify{-1}} \mathbf{a}(\tau, f) s_1(t-\tau,f),
    \label{eq:convolutive-model}
\end{align}
where $\mathbf{a}(0, f) \in \mathbb{C}^M$ is the relative transfer function (RTF) for the target source image, and $\mathbf{a}(\tau, f) \in \mathbb{C}^M$ is the convolutive RTF for the late reverberation.
In~\eqref{eq:instantaneous-model}, $s_1(t,f)$ is the target source image at the reference microphone, where we consider the reference microphone as the first channel without loss of generality.
In \eqref{eq:convolutive-model}, $\delta$ and $\Delta$ are the delay and tap length of the late reverberation, respectively.

The original WPD beamformer $\mathbf{w}(f) \in \mathbb{C}^{M(\Delta+1)}$ is obtained by solving the following optimization problem~\cite{Nakatani2019}:
\begin{equation}
    \min_{\mathbf{w}(f)} 
    \sum_{t=1}^T \frac{|\mathbf{w}^\mathsf{H}(f) \mathbf{y}(t,f)|^2}{\lambda(t,f)}
    \hspace{10pt}
    \mathrm{s.t.}
    \hspace{10pt}
    \mathbf{w}^\mathsf{H}(f) \mathbf{b}(f) = 1,
    \label{eq:wpd-original}
\end{equation}
where $\mathbf{y}(t,f) \triangleq [\mathbf{x}^\mathsf{T}(t,f), \mathbf{x}^\mathsf{T}(t-\delta, f), \ldots, \mathbf{x}^\mathsf{T}(t-\delta-\Delta+1, f)]^\mathsf{T} \in \mathbb{C}^{M(\Delta+1)}$,
$\mathbf{b}(f) \triangleq [\mathbf{a}^\mathsf{T}(0, f), 0, \ldots, 0]^\mathsf{T} \in \mathbb{C}^{M(\Delta+1)}$, $\lambda(t,f) \in \mathbb{R}_+$ is the power of the target signal, and \modify{$(\cdot)^\mathsf{H}$ denotes the Hermitian transpose}.
The original WPD beamformer in \eqref{eq:wpd-original} requires the RTF that is not easy to estimate by NN.
We thus use another formulation of the WPD beamformer~\cite{Zhang2020a} as the \texttt{SE} model in \eqref{eq:speech-enhancement}:
\begin{equation}
    \mathbf{w} (f) = \frac{\mathbf{R}^{-1}(f)\mathbf{H}(f)}{\mathrm{Trace}[\mathbf{R}^{-1}(f)\mathbf{H}(f)]} \mathbf{u},
    \label{eq:wpd-reformulated}
\end{equation}
where $\mathbf{u} = [1, 0\ldots, 0] \in \mathbb{C}^{M(\Delta+1)}$ is the one-hot vector denoting the reference microphone.
Here, $\mathbf{R}(f) \in \mathbb{C}^{M(\Delta+1) \times M(\Delta+1)}$ is the weighted multi-tap spatial covariance matrix computed using a time-frequency (T-F) mask $\mathcal{M}_m(t,f) \in [0, 1]$ as follows:
\begin{align}
\mathbf{R}(f) &= \sum_{t=1}^T \frac{\mathbf{y}(t,f) \mathbf{y}^\mathsf{H}(t,f)}{\lambda(t,f)}, \\
\lambda(t,f) &= \frac{1}{M} \sum_{m=1}^{M} \mathcal{M}_m(t,f) |x_m(t,f)|^2.
\label{eq:psd-estimation}
\end{align}
Meanwhile, $\mathbf{H}(f) \in \mathbb{C}^{M(\Delta+1) \times M(\Delta+1)}$ in \eqref{eq:wpd-reformulated} is the multi-tap spatial covariance matrix of the target source image:
\begin{align}
    \mathbf{H}(f) &= \frac{1}{\sum_{t=1}^T \underline{\mathcal{M}}(t,f)} \sum_{t=1}^T \underline{\mathcal{M}}(t,f) \underline{\mathbf{x}}(t,f) \underline{\mathbf{x}}^\mathsf{H}(t,f), \\
    \underline{\mathcal{M}}(t,f) &= \frac{1}{M} \sum_{m=1}^M \mathcal{M}_m(t,f),
    \label{eq:mask_averaging}
\end{align}
where $\underline{\mathbf{x}}(t,f) = [\mathbf{x}^\mathsf{T}(t,f), 0, \ldots, 0]^\mathsf{T} \in \mathbb{C}^{M(\Delta+1)}$.
The enhanced signal $\widehat{s}(t,f)$ is obtained by using the filter $\mathbf{w}(f)$ in \eqref{eq:wpd-reformulated} as:
\begin{equation}
    \widehat{s}(t,f) = \mathbf{w}^\mathsf{H}(f) \mathbf{y}(t,f).
    \label{eq:conv-beamforming}
\end{equation}
This result is converted to the time domain by inverse STFT (iSTFT).

The T-F mask $\mathcal{M}_m (t,f)$ in \eqref{eq:psd-estimation} and \eqref{eq:mask_averaging} is estimated by using NN from the STFT magnitude of the signal at the $m$th channel.
This NN is trained to minimize the convolutive transfer function invariant signal-to-distortion ratio (CI-SDR) loss~\cite{Boeddeker2021} of the enhanced time-domain signal.
The CI-SDR loss compensates for small signal shifts by using a short filter and makes the training more robust and stable.

Fully neural multi-channel SE methods have recently been developed in the time domain~\cite{Gu2019} and T-F domain~\cite{Tan2022}.
Although these methods demonstrated promising results in simulation, their performance often deteriorates on real recordings~\cite{Zhang2021}.
Meanwhile, mask-based beamforming is more robust to such a domain mismatch and can leverage an arbitrary number of microphones with different array geometry~\cite{Subramanian2019}.

\vspace{-0.2cm}
\subsection{Feature Extraction by WavLM}

While any SSLR model can be incorporated into our system, we propose to use WavLM~\cite{Chen2021} as it demonstrated its strength in the SUPERB benchmark tasks~\cite{Yang2021}.
WavLM consists of a convolutional encoder and multiple Transformer layers.
These are trained to predict the $k$-means cluster of the clean speech feature in the masked region from noisy and overlapped speech input.
This training is conducted with several datasets: Libri-light~\cite{Khan2020}, GigaSpeech~\cite{Chen2021a}, and VoxPopuli~\cite{Wang2021}.
We thus expect that WavLM can generalize to other datasets including CHiME-4 and REVERB challenge datasets.

In MultiIRIS, WavLM extracts the frame-wise feature from the enhanced time-domain signal, as in \eqref{eq:feature-extraction}.
Specifically, the features $\mathbf{Z}$ are obtained by a learnable weighted sum of the outputs from a convolutional encoder $\mathbf{Z}_0$ and Transformer layers $(\mathbf{Z}_1, \ldots, \mathbf{Z}_L)$:
\begin{equation}
\mathbf{Z} = \sum_{l=0}^L \alpha_l \mathbf{Z}_l, \label{eq:weighted-sum}
\end{equation}
where $L$ is the number of Transformer layers, and $\alpha_l \in [0, 1]$ is a learnable weight that satisfies $\sum_{l=0}^L \alpha_l = 1$.
This weight is optimized with the \texttt{ASR} model as explained in the next subsection.

\subsection{E2E ASR by \modify{Joint} CTC/Attention}

We adopt the \modify{joint} CTC/attention-based encoder--decoder~\cite{Watanabe2017} as the $\texttt{ASR}$ model.
It comprises an encoder, CTC, and a decoder as:
\begin{align}
    \mathbf{Q} &= \texttt{ConformerEnc}(\mathbf{Z}), \label{eq:asrenc}\\
    \mathbf{C}^{\text{(CTC)}} &= \texttt{CTC}(\mathbf{Q}), \label{eq:asrctc} \\
    \mathbf{C}^{\text{(Dec)}} &= \texttt{TransformerDec}(\mathbf{Q}), \label{eq:asrdec}
\end{align}
where $\mathbf{C}^{\text{(CTC)}}$ and $\mathbf{C}^{\text{(Dec)}}$ are the estimates from CTC and the decoder, respectively.
At the inference, we combine the posteriors from CTC and the decoder and apply a beam search.

The \texttt{ASR} model and the learnable weight $\alpha_l$ in \eqref{eq:weighted-sum} are optimized based on the following sum of two objective functions:
\begin{equation}
    \mathcal{L} = \beta \log p_{\mathrm{ctc}}(\mathbf{C} \mid \mathbf{Z})
    + (1-\beta) \log p_{\mathrm{att}}(\mathbf{C} \mid \mathbf{Z}), \label{eq:asr-loss}
\end{equation}
where $\beta \in [0, 1]$ is a hyperparameter, and \modify{$p_{\mathrm{ctc}}$ and $p_{\mathrm{att}}$ are the posterior distributions from CTC and the decoder, respectively.}
The CTC objective function uses the forward-backward algorithm in the training and enforces the alignment between the features and the transcription.
This will mitigate misalignments in the attention-based encoder--decoder.

\subsection{Training Procedure}

\modify{As joint training of \texttt{SE} and \texttt{ASR} models from scratch often results in suboptimal performance~\cite{Chang2022}, we separately train each model and fine-tine them together afterwards.}
In detail, we initially train the \texttt{SE} model using the CI-SDR loss with simulated multichannel data.
Meanwhile, the \texttt{ASR} model is trained based on \eqref{eq:asr-loss} with single-channel data.
Then, we jointly fine-tune these models to maximize the ASR performance with the \texttt{SSLR} model.
\modify{Without pre-training, the joint training was unstable and resulted in worse performance than the model without joint training.}
In this paper, we do not fine-tune the \texttt{SSLR} model, WavLM, because it is already trained on a large amount of data.
This can save the computational cost for fine-tuning and avoid overfitting to the training data.

\vspace{-4pt}
\section{Evaluation on CHiME4 Dataset}

We investigated SE and ASR performance on the CHiME-4 2- and 6-channel tracks~\cite{Vincent2017}.
Our experiments were conducted using the end-to-end speech processing (ESPnet) toolkit%
\footnote{The training scripts are available online: \url{https://github.com/espnet/espnet/tree/master/egs2/chime4/enh_asr1}.}%
~\cite{Watanabe2018,Lu2022}.

\begin{table}[t]
    \centering
    \caption{WER with different features on CHiME-4 dataset.
    BeamformIt was applied to observed noisy signal.}
    \scalebox{1.}[1.]{
    \begin{tabular}{c|l|cc|cc}
        \toprule
        \multicolumn{2}{c|}{} & \multicolumn{2}{c|}{Dev. Set} & \multicolumn{2}{c}{Test Set} \\
        \cmidrule{3-6}
        \multicolumn{2}{c|}{} & Simu. & Real & Simu. & Real \\
        \midrule
        \multirow{3}{*}{2ch.}
        & Fbank & 18.12 & 16.23 & 25.95 & 25.00 \\
        & HuBERT & 9.76 & 7.63 & 15.49 & 16.80 \\
        & WavLM & \bf{4.17} & \bf{5.33} & \bf{5.58} & \bf{4.57} \\
        \midrule
        \multirow{3}{*}{6ch.}
        & Fbank & 14.79 & 13.86 & 22.20 & 20.76 \\
        & HuBERT & 6.29 & 5.12 & 10.38 & 10.46 \\
        & WavLM & \bf{2.78} & \bf{4.28} & \bf{3.80} & \bf{3.57} \\
        \bottomrule
    \end{tabular}
    }
    \label{tab:chime4-sslr}
    \vspace{-8pt}
\end{table}

\vspace{-4pt}
\subsection{Dataset}

The CHiME-4 dataset provides real and simulated 6-channel noisy recordings \modify{at 16 kHz}~\cite{Vincent2017}.
The clean speech was from the Wall Street Journal (WSJ0) corpus, and the recordings contain four types of noise: bus, cafe, pedestrian, and street.
All channels except the second channel of the noisy recordings in the training set were used for the pre-training of the \texttt{ASR} model.
We also used clean speech in the WSJ0 and WSJ1 corpora based on a recipe in ESPnet~\cite{Watanabe2018}, as in \cite{Chang2022}.
The \texttt{SE} model was trained on the simulated recordings because this training requires pairs of clean and noisy signals.
Meanwhile, both simulated and real recordings were used in the joint training.

\begin{table}[t]
    \centering
    \caption{Enhancement performance on simulated test set.}
    \scalebox{0.97}[0.97]{
    \begin{tabular}{c|l|cccc}
        \toprule
        \multicolumn{2}{c|}{} & SDR & STOI & PESQ \\
        \midrule
        1ch. & Observed signal & 3.96 & 0.795 & 1.878 & \\
        \midrule
        \multirow{7}{*}{2ch.}
        & BeamformIt & 4.43 & 0.826 & 2.017 & \\
        \cmidrule{2-6}
        & MPDR & 9.87 & 0.879 & 2.255 \\
        & + Joint training & 9.87 & 0.880 & 2.254 \\
        \cmidrule{2-6}
        & MVDR & 10.42 & 0.882 & 2.282 \\
        & + Joint training & 10.40 & 0.882 & 2.282 \\
        \cmidrule{2-6}
        & WPD & \bf{11.02} & 0.879 & \bf{2.347} \\
        & + Joint training & 10.94 & \bf{0.886} & 2.338 \\
        \midrule
        \multirow{7}{*}{6ch.}
        & BeamformIt & 4.91 & 0.850 & 2.141 & \\
        \cmidrule{2-6}
        & MPDR & 14.20 & 0.939 & 2.505 \\
        & + Joint training & 14.17 & 0.939 & 2.506 \\
        \cmidrule{2-6}
        & MVDR & \bf{16.07} & \bf{0.952} & \bf{2.635} \\
        & + Joint training & 16.06 & 0.951 & \bf{2.635} \\
        \cmidrule{2-6}
        & WPD & \bf{16.07} & 0.949 & 2.630 \\
        & + Joint training & \bf{16.07} & 0.950 & 2.633 \\
        \bottomrule
    \end{tabular}
    }
    \label{tab:chime4-beamformer-enh}
    \vspace{-4pt}
\end{table}

\begin{figure}[t]
\centering
\vspace{-4pt}
\includegraphics[width=0.96\columnwidth]{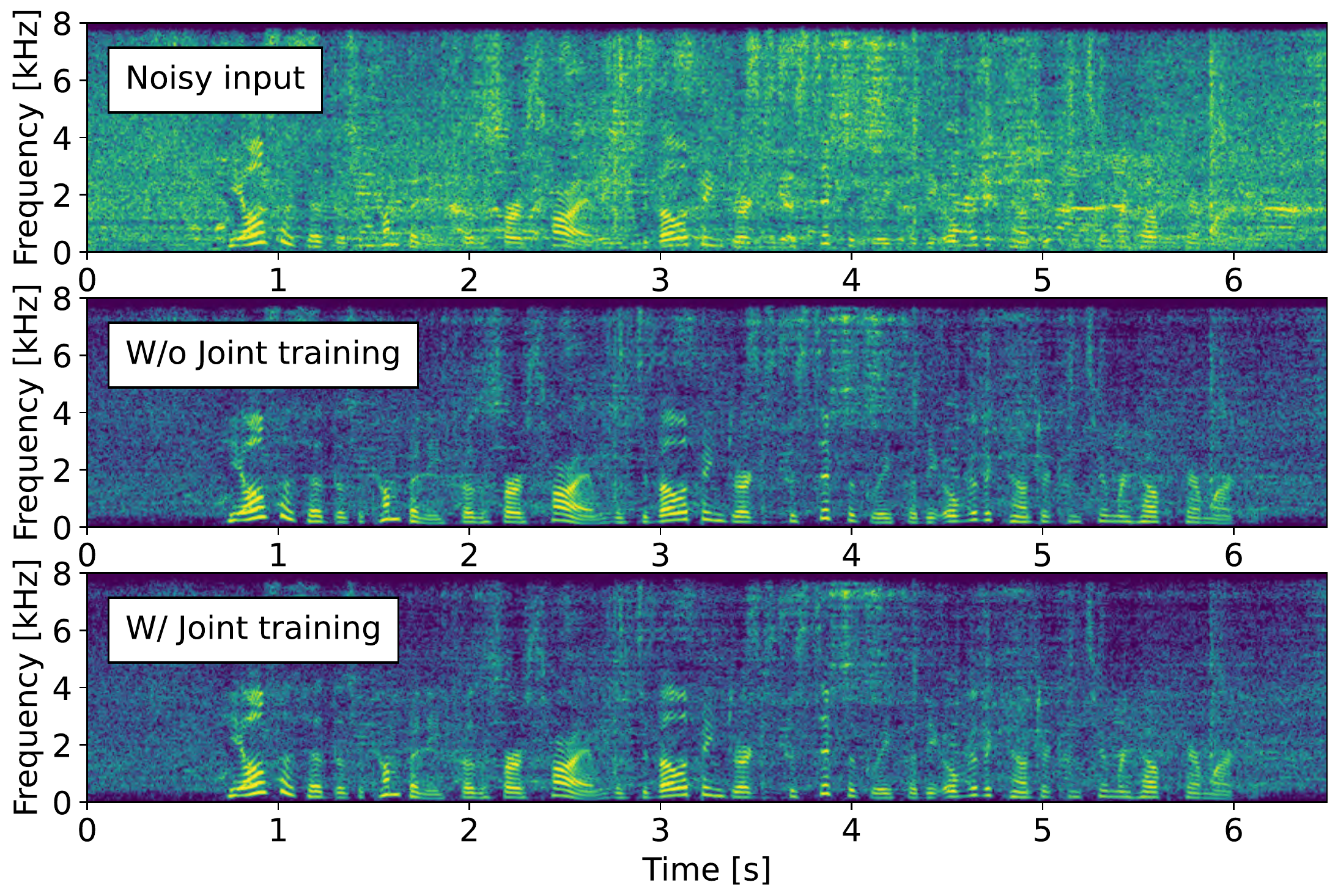}
\vspace{-8pt}
\caption{Examples of noisy observed signal and signals enhanced by the WPD beamformers. \modify{Pre-emphasis is applied to visualize high-frequency components more clearly.}
}
\label{fig:spectrogram}
\vspace{-10pt}
\end{figure}

\vspace{-4pt}
\subsection{Configurations}
\label{sec:config}

In the \texttt{SE} model, three-layered bidirectional long short-term memory (BLSTM) with a projection layer was used to estimate the T-F mask $\mathcal{M}_m(t,f)$ in \eqref{eq:psd-estimation} and \eqref{eq:mask_averaging}.
The BLSTMs had 512 units for both directions.
STFT was implemented with the Hann window of 400 samples with a 128-sample shift and 512 discrete Fourier transform points.
In the WPD beamformer, the number of delay $\delta$ in \eqref{eq:convolutive-model} was set to 3, and the tap length $\Delta$ was 5 and 3 for the 2- and 6-channel tracks, respectively.
The NN was optimized on the basis of the CI-SDR loss~\cite{Boeddeker2021} by using the Adam optimizer.
The initial learning rate was $4 \times 10^{-4}$, and it was halved when the loss on the development set was not improved in successive epochs.

In the \texttt{ASR} model, Conformer~\cite{Gulati2020} and Transformer~\cite{Karita2019} were used for the encoder and decoder, respectively.
\texttt{ConformerEnc} in \eqref{eq:asrenc} consisted of 12 layers where each layer had 4 attention heads with feed-forward layers of 2048 units.
The kernel size of the convolution layers was set to $15$.
For downsampling, additional convolutional layers were used.
\texttt{TransformerDec} in \eqref{eq:asrdec} consisted of 6 layers with 4 attention heads.
The dimensions of WavLM features $\mathbf{Z}$ in \eqref{eq:weighted-sum} were reduced from 1024 to 128 by a feed-forward layer.
The \texttt{ASR} model and the learnable weight $\alpha_l$ were optimized by using the Adam optimizer with the peak learning rate at $1 \times 10^{-3}$ and a warm-up of 20000 steps.
At the inference, we used a Transformer-based character-level language model.
We performed model averaging over the 10 checkpoints with the highest accuracy.
The joint training used the stochastic gradient descent (SGD) method with a learning rate of $4 \times 10^{-3}$ and momentum of $0.9$.
\modify{The SGD method is important to stabilize the fine-tuning and improve the performance.}

\begin{table}[t]
    \centering
    \caption{WER with different beamformers on CHiME-4 dataset.
    WavLM was used for feature extraction in all systems.}
    \scalebox{0.97}[0.97]{
    \begin{tabular}{c|l|cc|cc|c}
        \toprule
        \multicolumn{2}{c|}{} & \multicolumn{2}{c|}{Dev. Set} & \multicolumn{2}{c|}{Test Set} & \multirow{2}{*}{Ave.}\\
        \cmidrule{3-6}
        \multicolumn{2}{c|}{} & Simu. & Real & Simu. & Real & \\
        \midrule
        \multirow{7}{*}{2ch.}
        & BeamformIt & 4.17 & 5.33 & 5.58 & 4.57 & 4.89 \\
        \cmidrule{2-7}
        & MPDR & 2.53 & 2.03 & 2.26 & 2.98 & 2.43 \\
        & + Joint training & 2.45 & 1.93 & 2.19 & 2.89 & 2.35 \\
        \cmidrule{2-7}
        & MVDR & 2.38 & 2.13 & 2.11 & 3.14 & 2.41 \\
        & + Joint training & 2.30 & 1.98 & \bf{2.04} & 2.86 & 2.28 \\
        \cmidrule{2-7}
        & WPD & 2.28 & 2.06 & 2.30 & 3.63 & 2.52 \\
        & + Joint training & \bf{2.04} & \bf{1.66} & \bf{2.04} & \bf{2.65} & \bf{2.07} \\
        \midrule
        \multirow{7}{*}{6ch.}
        & BeamformIt & 2.78 & 4.28 & 3.80 & 3.57 & 3.60 \\
        \cmidrule{2-7}
        & MPDR & 1.36 & 1.44 & 1.39 & 1.84 & 1.49 \\
        & + Joint training & 1.36 & 1.42 & 1.36 & 1.79 & 1.47 \\
        \cmidrule{2-7}
        & MVDR & 1.21 & 1.38 & 1.23 & 1.91 & 1.41 \\
        & + Joint training & 1.25 & \bf{1.31} & \bf{1.21} & 1.85 & 1.39 \\
        \cmidrule{2-7}
        & WPD & \bf{1.19} & 1.32 & 1.29 & 1.85 & 1.39 \\
        & + Joint training & 1.22 & 1.33 & 1.24 & \bf{1.77} & \bf{1.38} \\
        \bottomrule
    \end{tabular}
    }
    \label{tab:chime4-beamformer-wer}
    \vspace{-10pt}
\end{table}

\vspace{-4pt}
\subsection{Effectiveness of Robust SSLR}

We evaluated the effectiveness of WavLM~\cite{Chen2021} with BeamformIt~\cite{Anguera2007} that is used in the CHiME-4 baseline.
We compared WavLM with log Mel-Filterbanks (Fbanks) with a 400-sample window and a 160-sample shift.
The number of bins was 80, and the delta and delta-delta coefficients were concatenated.
We also evaluated HuBERT~\cite{Hsu2021} trained on clean speech in Libri-light~\cite{Khan2020}.
These methods were implemented by the self-supervised speech pre-training and representation learning (S3PRL) toolkit~\cite{Yang2021}.

The results are summarized in Table~\ref{tab:chime4-sslr}.
We can observe that, by using HuBERT and WavLM, the word error rate (WER) was improved from that of Fbank.
This result shows the strength of SSLR models pre-trained on a large amount of unlabeled data.
WavLM outperformed HuBERT on both simulated and real datasets, and thus SSL with noisy and overlapped data is effective.

\vspace{-4pt}
\subsection{Effectiveness of MultiIRIS on CHiME-4 Datasets}
\label{sec:different-beamformers}

\modify{To demonstrate the effectiveness of the WPD beamformer, we compared it with the MPDR and MVDR beamformers.}                    
As explained in Section~\ref{sec:mcse_for_asr}, the MPDR and MVDR beamformers focus on denoising while the WPD beamformer jointly performs dereverberation and denoising by convolutional beamforming in \eqref{eq:conv-beamforming}.
These beamformers were also computed based on the T-F mask estimated by the NN described in Section~\ref{sec:config}
In addition to WER, we evaluated the enhancement performance by using the signal-to-distortion ratio (SDR), the short-time objective intelligibility (STOI), and the perceptual evaluation of speech quality score (PESQ).

Table~\ref{tab:chime4-beamformer-enh} shows the enhancement performance.
\modify{While our joint training did not use the CI-SDR loss, it maintained high enhancement performance contrary to a previous work~\cite{Subramanian2019}.}
We can also observe this in  Fig.~\ref{fig:spectrogram} where we report the STFT magnitude of the signals enhanced by the WPD beamformer before and after joint training\footnote{
{Additional STFT magnitude and audio examples are available online: \url{https://popcornell.github.io/MultiIRIS-demo}.}
}.
This is a strong advantage as the front-end, even after fine-tuning, can also be used for tasks other than ASR.

WERs with different beamformers are summarized in Table~\ref{tab:chime4-beamformer-wer}.
All the mask-based beamformers improved WER compared with BeamformIt.
That is, strong multi-channel SE methods are important for ASR even when a robust SSLR is used.
As a result of unifying WPE and the MPDR beamformer, the WPD beamformer improved WER.
While MVDR also worked well, the WPD beamformer with joint training performed best on average.
\modify{Regarding feature extraction, the last layer was the most significant ($\alpha_{24} = 0.8$).
A similar tendency was observed with HuBERT in \cite{Chang2021}.
}

\begin{table}[t]
    \centering
    \caption{Comparison with existing systems on CHiME-4 dataset.
    Note that IRIS and MultiIRIS used external data to train WavLM.}
    \scalebox{1.}[1.]{
    \begin{tabular}{c|l|cc|cc}
        \toprule
        \multicolumn{2}{c|}{} & \multicolumn{2}{c|}{Dev. Set} & \multicolumn{2}{c}{Test Set} \\
        \cmidrule{3-6}
        \multicolumn{2}{c|}{} & Simu. & Real & Simu. & Real \\
        \midrule
        \multirow{2}{*}{1ch.}
        & Kaldi baseline~\cite{Chen2018} & 6.81 & 5.58 & 12.15 & 11.42 \\
        & IRIS~\cite{Chang2022} & 3.16 & 2.03 & 6.12 & 3.92 \\
        \midrule
        \multirow{4}{*}{2ch.}
        & Kaldi baseline~\cite{Chen2018} & 3.94 & 2.85 & 5.03 & 5.40 \\
        & Du \emph{et al.}~\cite{Du2016} & 3.46 & 2.33 & 5.74 & 3.91 \\
        & Wang \emph{et al.}~\cite{Wang2020} & 2.17 & 1.99 & 2.53 & 3.19 \\
        \cmidrule{2-6}
        & MultiIRIS & \bf{2.04} & \bf{1.66} & \bf{2.04} & \bf{2.65} \\
        \midrule
        \multirow{4}{*}{6ch.}
        & Kaldi baseline~\cite{Chen2018} & 2.10 & 1.90 & 2.66 & 2.74 \\
        & Du \emph{et al.}~\cite{Du2016} & 1.78 & 1.69 & 2.12 & 2.24 \\
        & Wang \emph{et al.}~\cite{Wang2020} & \bf{1.15} & 1.50 & 1.45 & 1.99 \\
        \cmidrule{2-6}
        & MultiIRIS & 1.22 & \bf{1.33} & \bf{1.24} & \bf{1.77} \\
        \bottomrule
    \end{tabular}
    }
    \label{tab:exp-comparison}
    \vspace{-8pt}
\end{table}

\vspace{-4pt}
\subsection{Comparison with Existing Methods on CHiME-4 Dataset}

Finally, we compared MultiIRIS with existing systems.
We would like to stress that WavLM used additional data in the pre-training stage, and thus this is not a fair comparison according to the CHiME-4 Challenge rules. 
This comparison, however, can provide an insight over ASR in noisy reverberant conditions when external resources are leveraged.
In Table~\ref{tab:exp-comparison}, we summarize the performance of existing systems and MultiIRIS with the WPD beamformer.
MultiIRIS consistently outperformed the single-channel IRIS~\cite{Chang2022} which also uses WavLM as the \texttt{SSLR} model. 
Hence, multi-channel SE can bring substantial benefits over single-channel SE, even with a strong back-end based on robust SSLR.
In both 2- and 6-channel tracks, MultiIRIS substantially outperformed the first ranking system~\cite{Du2016} and the state-of-the-art system~\cite{Wang2020}.
\modify{When WavLM was replaced by HuBERT, WERs for MultiIRIS were $2.33\%$ and $4.66\%$ on the simulated and real test sets.
This is because the pre-training of HuBERT uses only clean speech.
}

\vspace{-4pt}
\section{REVERB Challenge}
\vspace{-4pt}

We also validated MultiIRIS on the REVERB dataset~\cite{Kinoshita2016}.
The training configuration was the same as the previous experiment.

\vspace{-4pt}
\subsection{Dataset}

We used clean speech in WSJ0, WSJ1, and WSJCAM0 corpora to pre-train the \texttt{ASR} model in accordance with a recipe in ESPnet~\cite{Watanabe2018}.
The \texttt{SE} model was trained on 8-channel simulated noisy reverberant recordings from the REVERB training set.
Then, the joint training was conducted on the same dataset.
The REVERB development and test sets contain both real and simulated recordings.

\vspace{-4pt}
\subsection{Effectiveness of MultiIRIS on REVERB Datasets}

As we confirmed the effectiveness of the WPD beamformer in Section~\ref{sec:different-beamformers}, we compared MultiIRIS with the WPD beamformer and our baseline that used WPE and BeamformIt (the same front-end used in the REVERB challenge baseline).
In our baseline, Fbank features were passed to the \texttt{ASR} model.
Table~\ref{tab:reverb_ablation} shows WER on the REVERB development and test sets.
MultiIRIS clearly outperformed our baseline (WPE + BeamformIt + Fbank).
Although the pre-training of WavLM does not take care of reverberation explicitly, MultiIRIS achieved promising results thanks to the end-to-end integration with the WPD beamformer.

Next, we compared MultiIRIS with existing systems.
As we used additional data to train E2E ASR and WavLM, this is also not a fair comparison.
We also evaluated MultiIRIS trained on the CHiME-4 dataset to investigate the generalization capability of our end-to-end integration.
In Table~\ref{tab:reverb_comparison}, MultiIRIS trained on the CHiME-4 and REVERB datasets outperformed the state-of-the-art system for the REVERB challenge 8-channel track~\cite{Wang2020a}.
Hence, our end-to-end integration of multi-channel SE with SSLR can also bring substantial improvements for ASR under reverberant conditions.
The performance of MultiIRIS trained on the CHiME-4 dataset was limited compared to that trained on the REVERB dataset.
This degradation could come from the longer reverberation of the real recordings in the REVERB dataset compared to the CHiME-4 dataset.

\begin{table}[t]
    \centering
    \caption{WER on REVERB challenge 8-channel track.}
    \scalebox{1.}[1.]{
    \begin{tabular}{l|cc|cc}
        \toprule
        \multirow{2}{*}{} & \multicolumn{2}{c|}{Dev. Set} & \multicolumn{2}{c}{Test Set} \\
        \cmidrule{2-5}
         & Near & Far & Near & Far \\
        \midrule
        \multicolumn{5}{c}{\emph{Simu.}} \\
        \midrule
        WPE + BeamformIt + Fbank & 2.9 & 3.8 & 3.5 & 4.2 \\
        MultiIRIS & \bf{0.9} & \bf{1.0} & \bf{1.2} & \bf{1.3} \\
        \midrule
        \multicolumn{5}{c}{\emph{Real}} \\
        \midrule
        WPE + BeamformIt + Fbank & 8.1 & 10.0 & 6.3 & 7.7 \\
        MultiIRIS & \bf{1.2} & \bf{1.6} & \bf{0.9} & \bf{1.1} \\
        \bottomrule
    \end{tabular}
    }
    \label{tab:reverb_ablation}
\end{table}

\begin{table}[t]
    \centering
    \caption{Comparison with existing systems on REVERB real set.
    Note that MultiIRIS used additional data to train WavLM.}
    \scalebox{1.}[1.]{
    \begin{tabular}{l|cc|cc}
        \toprule
        \multirow{2}{*}{} & \multicolumn{2}{c|}{Dev. Set} & \multicolumn{2}{c}{Test Set} \\
        \cmidrule{2-5}
         & Near & Far & Near & Far \\
        \midrule
        Kaldi baseline~\cite{Kinoshita2016} & - & - & 50.1 & 47.6 \\
        Delcroix \emph{et al.}~\cite{Delcroix2015} & - & -  & 8.9 & 9.3 \\
        Wang \emph{et al.}~\cite{Wang2020a} & 7.9 & 7.7 & 5.9 & 6.4 \\
        \midrule
        MultiIRIS (CHiME-4) & 6.4 & 7.7 & 4.6 & 6.1 \\
        MultiIRIS (REVERB) & \bf{1.2} & \bf{1.6} & \bf{0.9} & \bf{1.1} \\
        \bottomrule
    \end{tabular}
    }
    \label{tab:reverb_comparison}
\end{table}

\vspace{-5pt}
\section{Conclusion}

In this paper, we propose an end-to-end integration of the WPD beamformer, WavLM, and E2E ASR for ASR under noisy reverberant conditions.
The \texttt{SE} and \texttt{ASR} models are pre-trained separately, and then these models are entirely fine-tuned with the \texttt{SSLR} model.
Our system outperformed existing systems on CHiME-4 and REVERB datasets by combining multi-channel speech enhancement with WavLM pre-trained on a large amount of additional data.
This result highlights the efficacy of combining SSLR with beamforming techniques for ASR under noisy reverberant conditions.

\vspace{-0.2cm}
\section{Acknowledgements}
\vspace{-0.2cm}

This work was supported by JSPS KAKENHI Grant Numbers JP21J21371, and JST CREST Grant Number JPMJCR19A3, Japan.

\clearpage
\section{References}
\begingroup
\setstretch{0.78}
\setlength\bibitemsep{0.85pt}
\printbibliography
\endgroup

\end{document}